\documentclass[conference]{IEEEtran}

\usepackage{graphicx}
\usepackage{svg}
\usepackage{subcaption}
\usepackage{hyperref} 
\usepackage{amsmath} 
\usepackage{amssymb} 
\usepackage{fancyhdr} 
\usepackage{pifont} 
\usepackage{xcolor}
\usepackage{soul}
\usepackage{adjustbox}
\usepackage{pgfgantt}
\usepackage{cite}
\usepackage{upgreek}
\usepackage{physics}
\ifCLASSINFOpdf
\else
\fi

\IEEEoverridecommandlockouts

\begin{document}

\title{Modeling Quantum Links for the Exploration of Distributed Quantum Computing Systems}

\author{
    Sahar Ben Rached$^{\dagger}$,
    Zezhou Sun$^{\ddagger}$,
    Junaid Khan$^{\dagger}$,
    Guilu Long$^{\ddagger}$,
    Santiago Rodrigo$^{\dagger}$,\\
    Carmen G. Almudéver$^{\S}$,
    Eduard Alarcón$^{\dagger}$,
    Sergi Abadal$^{\dagger}$\\[1ex]
    $^{\dagger}$NanoNetworking Center in Catalunya, Universitat Politècnica de Catalunya, Barcelona, Spain\\
    $^{\ddagger}$Department of Physics, Tsinghua University, Beijing, China\\
    $^{\S}$Computer Engineering Department, Universitat Politècnica de València, València, Spain\thanks{Authors gratefully acknowledge funding from the European Commission through HORIZON-EIC-2022-PATHFINDEROPEN-01-101099697 (QUADRATURE) and HORIZON-ERC-2021-101042080 (WINC), from the QCOMM-CAT-Planes Complementarios: Comunicacion Cuántica - supported by MICIN with funding from the European Union, NextGenerationEU (PRTR-C17.I1), by Generalitat de Catalunya, by ICREA Academia Award 2024, and from the Spanish Ministry of Science, Innovation and Universities through the Beatriz Galindo program (BG20-00023).}
    }

\markboth{Journal of \LaTeX\ Class Files,~Vol.~14, No.~8, August~2015}%
{Shell \MakeLowercase{\textit{et al.}}: Bare Demo of IEEEtran.cls for IEEE Journals}

\maketitle
\begin{abstract}
Quantum computing offers the potential to solve certain complex problems, yet, scaling monolithic processors remains a major challenge. Modular and distributed architectures are proposed to build large-scale quantum systems while bringing the security advantages of quantum communication. At present, this requires accurate and computationally efficient models of quantum links across different scales to advance system design and guide experimental prototyping. In this work, we review protocols and models for estimating latency, losses, and fidelity in quantum communication primitives relying on quantum state distribution via microwave photons. We also propose a scalable simulation framework to support the design and evaluation of future distributed quantum computing systems.
\end{abstract}


\IEEEpeerreviewmaketitle

\section{Introduction}
\label{sec:introduction}
Quantum computing is an emerging computational paradigm aimed at addressing problems intractable for classical computers, such as integer factorization\cite{shor1999polynomial}, quantum system simulation\cite{daley2022practical}, and combinatorial optimization \cite{abbas2024challenges}. However, achieving the full potential of quantum computing requires scaling from the currently available capacity and size limited processors\cite{abughanem2025ibm}, to tens of thousands or millions of qubits and supporting error correction protocols.

\begin{figure}
\centering
\begin{subfigure}{0.5\columnwidth}
  \centering
  \includegraphics[width=1\textwidth]{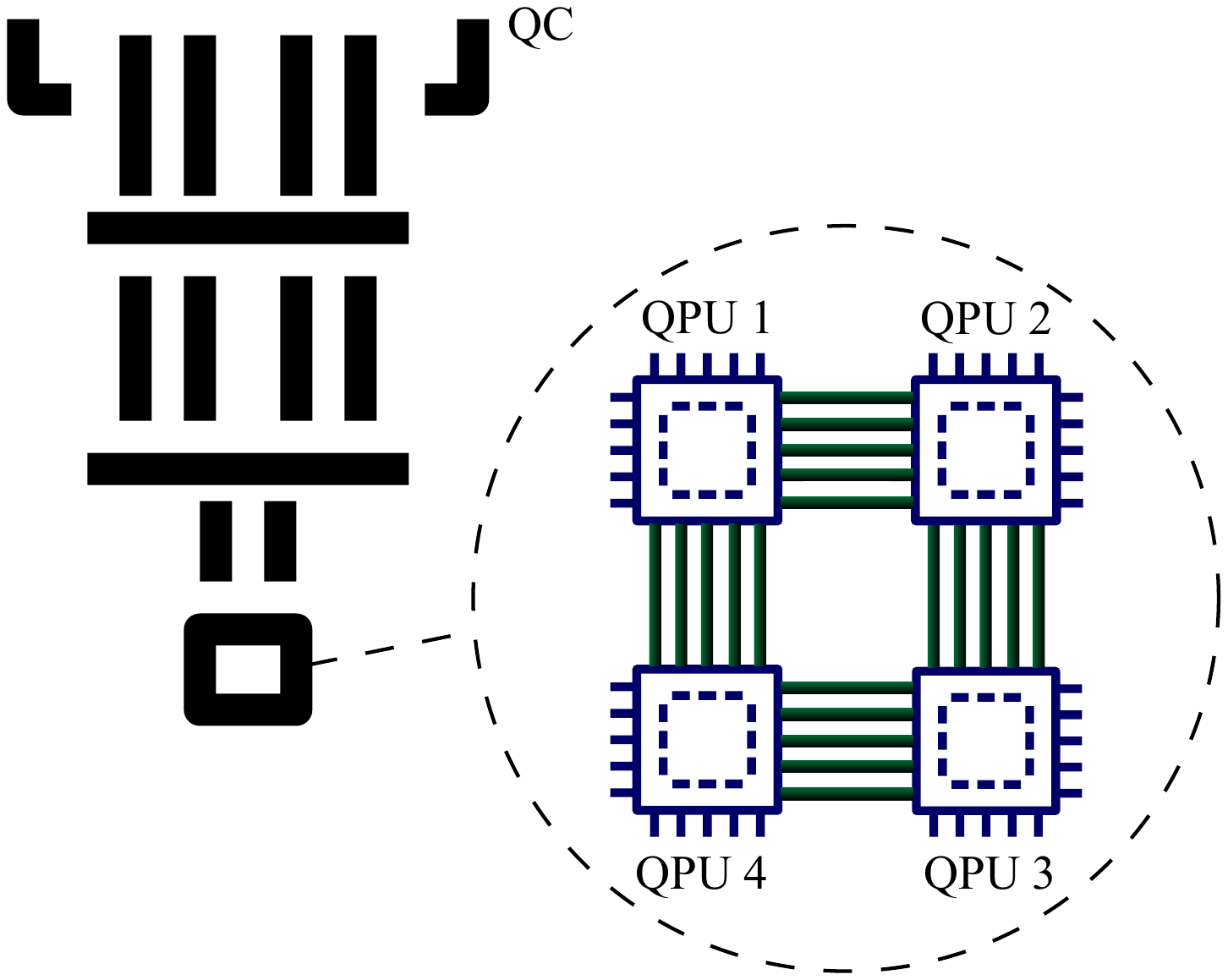}
  \caption{}
  \label{modular_qc}
\end{subfigure}%
\begin{subfigure}{0.5\columnwidth}
  \centering
  \includegraphics[width=1\textwidth]{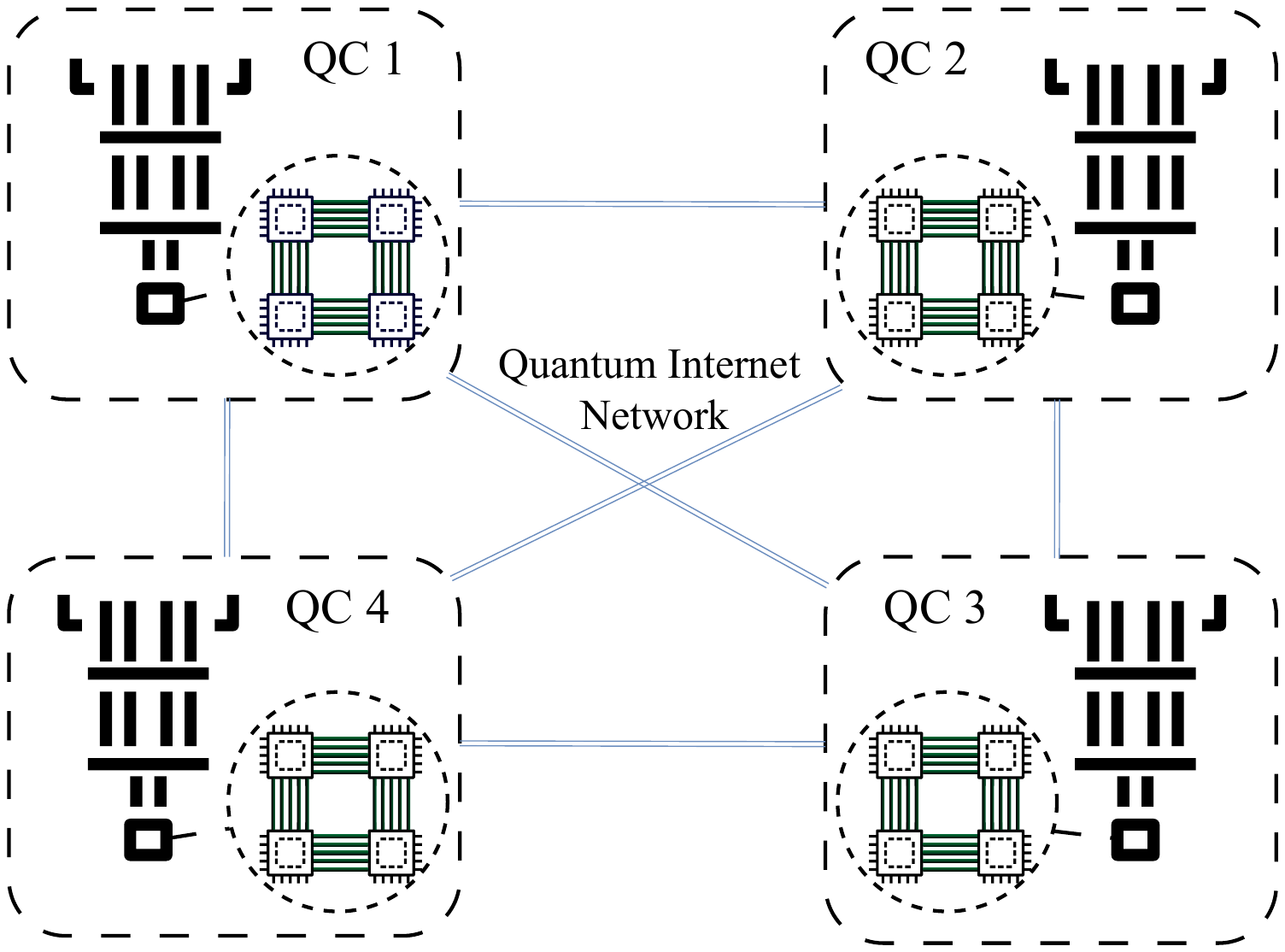}
  \caption{}
  \label{distributed_qc}
\end{subfigure}%
\vspace{-0.1cm}
\caption{Illustration of (a) a modular quantum processor installed within a cryogenic system, where quantum processing units (QPUs) are interconnected through cavity-mediated inter-core links, and (b) a distributed quantum computing architecture, where modular processors installed in separate cryogenic systems are interconnected via a large-scale quantum network.}
\label{qc_systems} \vspace{-0.3cm}
\end{figure}

Modular and distributed quantum architectures provide a promising path for scaling current systems\cite{bravyi2022future}. Modular quantum computing architectures, as illustrated in Fig.~\ref{modular_qc}, are based upon connecting smaller-sized quantum chips via quantum communication networks to enable cross-chip state transmission, entanglement generation \cite{ferrari2023modular}, or direct quantum operations \cite{norris2025performance} using short-range cavities or waveguides. One of the major experimental methods to establish quantum networks for modular systems involves cavity-based links \cite{ramette2022any}. Alternative configurations include entanglement generation networks, particularly for Bell pair distribution between targeted qubits \cite{matsukevich2006entanglement}, and infrastructures that support key quantum communication primitives such as the teleportation protocol \cite{doi:10.1126/science.1253512}. Another promising direction is the development of transduction-based networks \cite{weaver2024integrated}, which enable long-range, high-fidelity state transmission via microwave-to-optical conversion.

Distributed quantum computing, depicted in Fig.~\ref{distributed_qc}, aims to interconnect multiple quantum computing systems located at spatially separated nodes via the quantum internet \cite{cuomo2020towards}. This approach is envisioned to scale the capacity of quantum computing systems while leveraging the security advantages of quantum communication. It is expected that the quantum internet will rely heavily on optical fiber infrastructure \cite{chung2021illinois}; however, its operation differs fundamentally from classical networking. Novel communication protocols must be developed to adhere to the principles of quantum information theory, entanglement-based information transfer~\cite{10612809}, and the integration of quantum repeaters for preserving coherence over long distances \cite{cacciapuoti2019quantum}.

Current research on modular and distributed quantum computers remains mainly in the prototype phase, focusing on experimental demonstrations. To support the early-stage exploration of such systems and optimal specification and design of their links, methods for fast yet accurate modeling are needed. 

In this context, this paper aims at reviewing and illustrating methods for the modeling and simulation of quantum-coherent communications, usable for both integrated links for modular architectures and fiber-based links for distributed quantum computers. 

In particular, the main contribution of our work is threefold. First, we tackle the analytical modeling of arbitrary state transmission in cavity-based links incorporating the cavity decay rate $\kappa$, qubit decay rate $\gamma$, and qubit-cavity coupling strength $g$ as main modeling parameters. Second, we also provide an analytical model of the Stimulated Raman Adiabatic Passage (STIRAP) protocol aiming to enhance the fidelity of cross-chip operations. Third, we illustrate both models to analyze the performance of links for arbitrary states as a function of link parameters, number of hops, or the protocol used.

\section{Background and Related Works}
\label{sec:background}
Building modular and distributed quantum computing systems depends largely on the accurate modeling and performance evaluation of quantum links interconnecting modules. In this light, research on coherent and deterministic qubit-to-qubit state transmission is of great significance~\cite{10181857}. 

Within the quantum computing context, a qubit state coupled to one or more interconnected mediator modes (where the mediator may be a cavity, waveguide, or fiber) enables its transmission coherently and deterministically over an extended distance to an end node, to be then absorbed by a target qubit. 

The commonly used model to describe such atom-cavity systems is the Jaynes-Cummings (JC) model~\cite{10.1088/978-0-7503-3447-1}, characterizing the interaction between a two-level system and a quantized electromagnetic field mode, and accurately describing the coupling process of light and matter according to quantum mechanics laws. This system is subject to dissipation and decoherence effect due to interactions with the environment, inducing loss of quantum states. Multiple methods have been proposed to describe open quantum systems, such as the Lindblad master equation~\cite{PhysRevApplied.22.024006}. Subsequently, point-to-point channel models using different mediators are studied using different encoding or transmission protocols~\cite{johnston2024cavity}. 

To improve the fidelity of state transfer, STIRAP is proposed~\cite{Bergmann_2019}. This protocol suggests to change the coupling strength between the atoms and the cavity on both ends to improve the fidelity. Several works have experimentally achieved high-fidelity quantum state transmission using STIRAP~\cite{PhysRevA.98.053413}.

\section{Methodology}
\label{sec:methodology}
\subsection{Qubit-Waveguide Quantum Model Dynamics}
\begin{figure}
\centering
\includegraphics[width=\columnwidth]{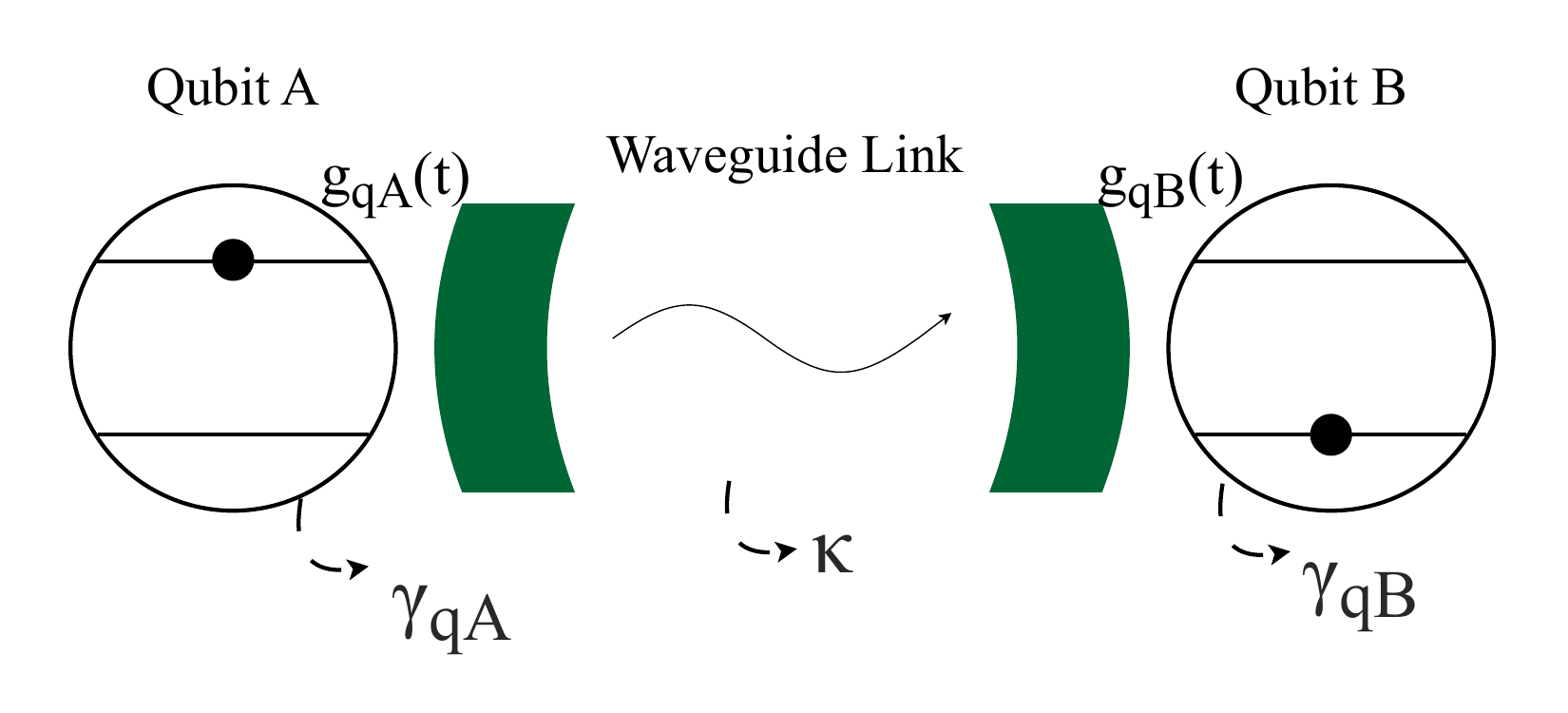} 
\caption{An illustration of the qubit-waveguide system dynamics as described with the quantum electrodynamics formalism.}
\label{qubit-cavity}
\end{figure}

In our model, we consider two spatially separated qubits, A and B, initialized in the ground and excited states, respectively, interacting via a shared waveguide-based communication channel as illustrated in Fig.~\ref{qubit-cavity}. The Hilbert space of the combined system is defined as the tensor product of the individual Hilbert spaces of the subsystem ($\mathcal{H} = \mathcal{H}_\text{waveguide} \otimes \mathcal{H}_{\text{qubit A}} \otimes \mathcal{H}_{\text{qubit B}}$), spanned by $2^3 = 8$ basis states represented as $\ket{n_A, n_W, n_B}$, where $n_i \in \{0, 1\}$. Initially at $t=0$, we prepare state vector $|\psi(0)\rangle = |0,1,0\rangle$, with qubit A excited, whereas the waveguide and qubit B remain in their ground states.
The described system dynamics are governed by the following Hamiltonian in the rotating-wave approximation (RWA) expressed as:
\begin{align}
    H &= \hbar \omega_q (\sigma_A^+\sigma_A^- + \sigma_B^+\sigma_B^-) + \hbar \omega_w a^\dagger a \nonumber \\
    &\quad + \hbar g_{qw}\left(\sigma_A^+ a + \sigma_A^- a^\dagger + \sigma_B^+ a + \sigma_B^- a^\dagger\right),
\end{align}
where $\omega_q$ and $\omega_w$ are the qubit and waveguide resonance frequencies, and $g_{qw}$ represents the coupling strength between the qubits and the waveguide, assumed symmetric in this work ($g_{qw} = g_{qA} = g_{qB}$), and that can be time-variant. The operators $\sigma^\pm$ are the Pauli raising/lowering operators for the qubits, while $a$ and $a^\dagger$ denote annihilation and creation operators for waveguide photon mode.

In an ideal closed system, the time evolution of the pure state $\ket{\psi(t)}$ follows the Schrödinger equation $\ket{\psi(t)} = e^{-iHt/\hbar}\ket{\psi(0)}$. However, realistic quantum systems interact with their environment, causing decoherence and dissipation. We model these effects using the Lindblad master equation, which describes the evolution of the system's density matrix over time $\rho(t)$ as 
\begin{equation}
    \begin{aligned}
     \frac{d\rho}{dt}& = -\frac{i}{\hbar}[H, \rho] + \sum_{j=A,B}\gamma_{qj} \left( \hat{\sigma}_j \rho \hat{\sigma}_j^\dagger - \frac{1}{2} \{\hat{\sigma}_j^\dagger \hat{\sigma}_j, \rho\} \right) + \\
       &+ \kappa \left( \hat{a} \rho \hat{a}^\dagger - \frac{1}{2} \{\hat{a}^\dagger \hat{a}, \rho\} \right), 
    \end{aligned}
\end{equation}
where the first term represents the coherent evolution under $H$, while the qubit energy decay ($\gamma_{qj}$) and waveguide photon loss ($\kappa$) are the Lindblad operators describing dissipative dynamics.

\subsection{STIRAP method}
STIRAP is used to achieve low-loss quantum state transfer. It adjusts the coupling coefficient between the intermediate medium system and the atoms. In principle, we first act on the coupling factor $g_{qB}$ between the target qubit state and the intermediate state, and then we act on the coupling factor $g_{qA}$ between the initial qubit state and the intermediate state. These actions overlap so that the system is always in a \emph{dark state}, thereby avoiding the occupation of the intermediate state and leading to reducing losses and improving the fidelity of state transmission. Depending on the type of system, time variation of the coupling coefficients $g_{qA}$ and $g_{qB}$ can be achieved by adjusting the pump pulse intensity or the coupler frequency.

In this paper, we use Gaussian functions to simulate the time domain envelopes of the two coupling coefficients $g_{qA}$ and $g_{qB}$, which are in the form of
\begin{align}
  \label{eq:2}
 g _A\left ( t \right ) =g _{0A} e^{-\frac{\left ( t-t_{delay} \right ) ^2}{T^2} } ,\\
 g _B\left ( t \right ) =g _{0B} e^{-\frac{t ^2}{T^2} } ,
\end{align}
where $g_0$, $t_{delay}$, and $T$ represent the maximum coupling coefficient amplitude, the relative delay between the two pulses, and the pulse width, respectively. These parameters can be adjusted to select the waveform that yields the best state transmission fidelity rate.

\subsection{Quantum Information Metrics}
To quantify the efficiency of quantum state transfer, we measure the received state fidelity and quantum coherent information.

We assume that the initial state of qubit A at \( t = 0 \) is
$\rho_A(0) = \cos\left (   \theta /2\right ) \left | 0  \right \rangle +e^{i\phi } \sin \left (   \theta /2\right )\left | 1  \right \rangle$, and the initial state of qubit B is $\rho_B(0) = |0\rangle$. Here, $\theta\in \left ( 0,\pi \right )  ,\ \phi \in \left ( 0,2\pi \right )$, meaning that we can transmit an arbitrary qubit state. The transmitted state through the cavity or fiber, received at qubit B, is $\rho_{Bd} =\cos\left (   \theta /2\right ) \left | 0  \right \rangle +e^{i\phi } \sin \left (   \theta /2\right )\left | 1  \right \rangle$.
We define the fidelity of the transmission during the evolution process as:
\[
F = \text{Tr}\left[\rho_{Bd} \rho_B(t)\right].
\]
The quantum coherent information, analogous to classical mutual information, quantifies how effectively quantum correlations are preserved during transmission, expressed as:
\begin{equation}
I(\rho,\varepsilon) = S(\varepsilon(\rho)) - S(\rho,\varepsilon),
\end{equation}
where $\rho$ represents the quantum information to be transmitted and $\varepsilon$ represents the quantum channel used to transmit it. $S(\rho,\varepsilon)$ is the Von Neumann entropy exchange induced by the quantum operation $\varepsilon$ upon the state $\rho$. 

\section{Results}
\label{sec:results}
\begin{figure}[!t]
  \centering
  \subfloat[]{\includegraphics[width=4.3 cm]{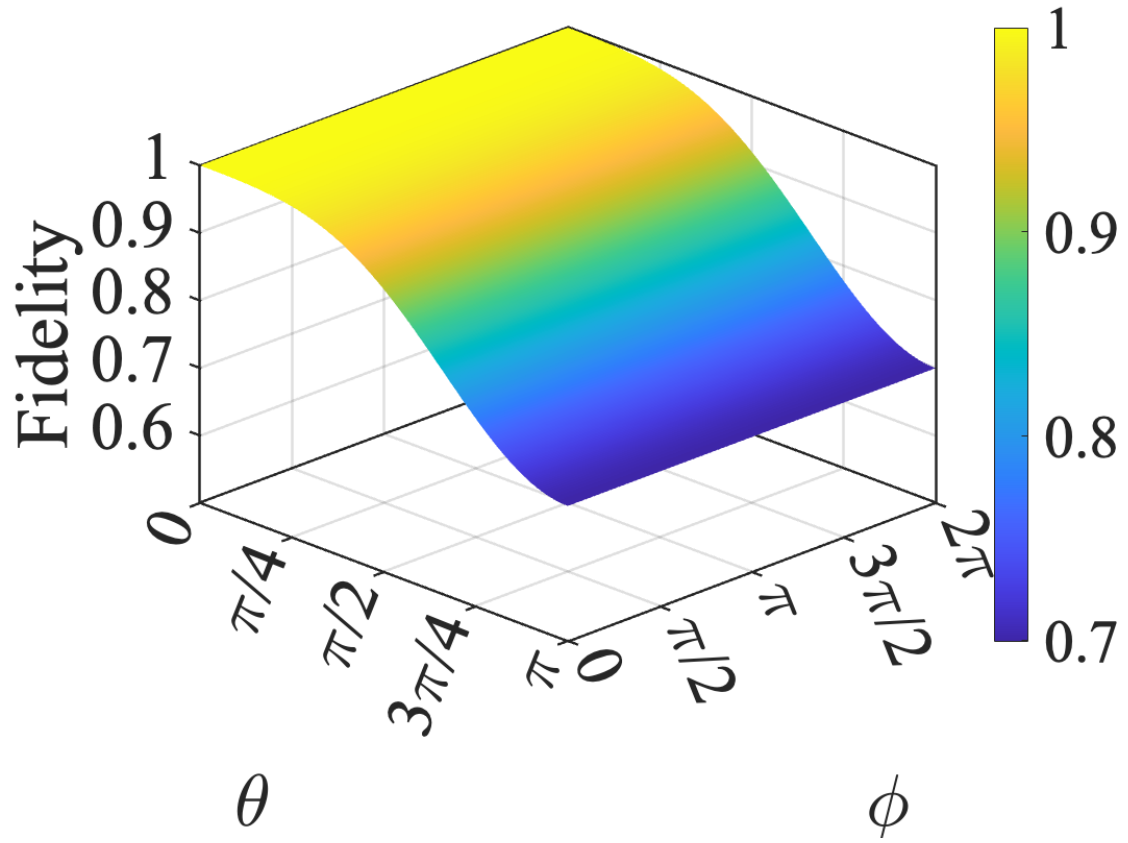}%
  \label{Fig:4a}}
  \hfil
  \subfloat[]{\includegraphics[width=4.3 cm]{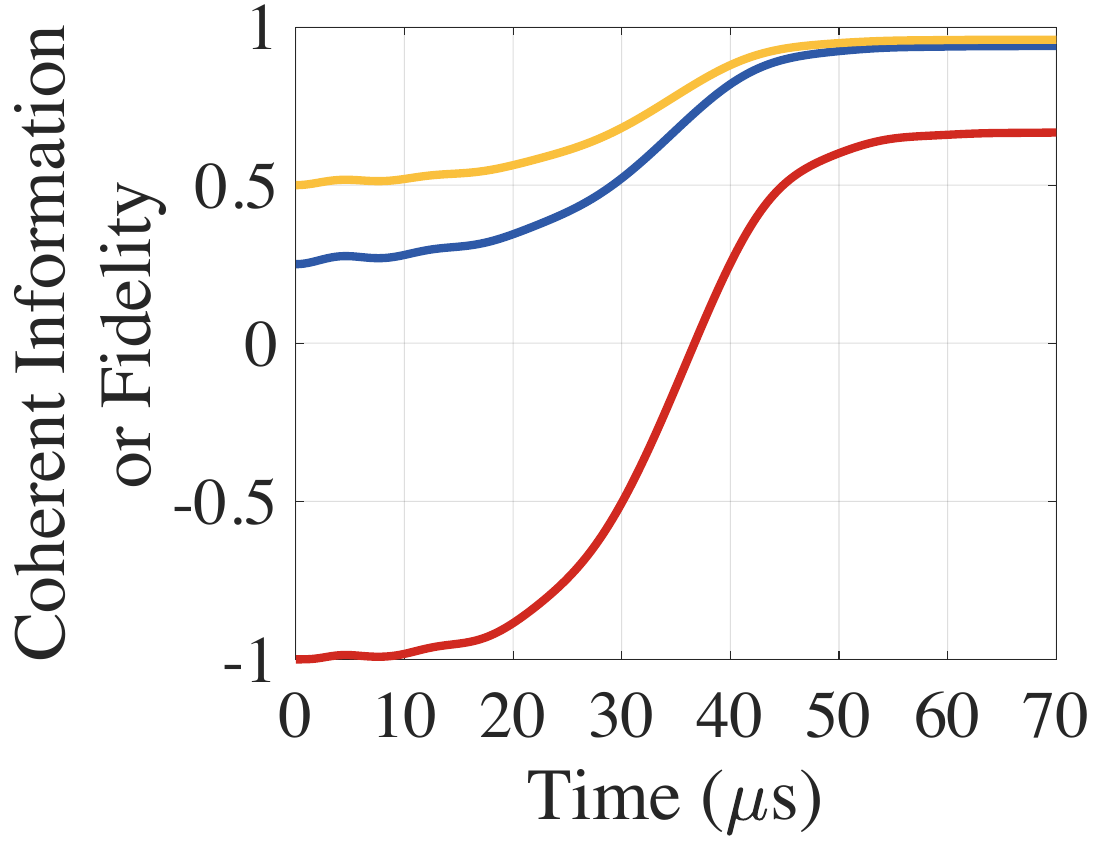}%
  \label{Fig:4b}}
  \caption{\label{fig:4} (a) Transmission fidelity of an arbitrary quantum state $\rho = \cos\left (\theta /2\right ) \left | 0  \right \rangle +e^{i\phi} \sin \left (   \theta /2\right )\left | 1  \right \rangle$ via cavity-based channels, with state-of-the-art parameters $g_0=5.8 \times 2\pi\ \rm{MHz}$, $\kappa=0.34 \times 2\pi\ \rm{MHz}$, and $\gamma=6 \times 2\pi\ \rm{MHz}$~\cite{PRXQuantum.3.010344}. (b) Comparison of quantum coherent information and fidelity for transmitting quantum states via a cavity-based channel. The red, blue, and yellow curves represent quantum coherence information, entanglement fidelity, and random single qubit fidelity, respectively, with $\kappa=0.04 \times 2\pi\ \rm{MHz}$. }
\end{figure}

\subsection{Arbitrary Qubit Transmission Performance}
In Fig.~\ref{fig:4}, we present the fidelity and quantum coherence information of transmitting arbitrary quantum states via a cavity-based channel. The simulation results of quantum coherent information indicates that our model can not only obtain highly-fidelity quantum states, but also validate the transmission of quantum information.

\subsection{State Transmission with and without STIRAP} 
\begin{figure}[!t]
  \centering
  \subfloat[]{\includegraphics[width=4.3 cm]{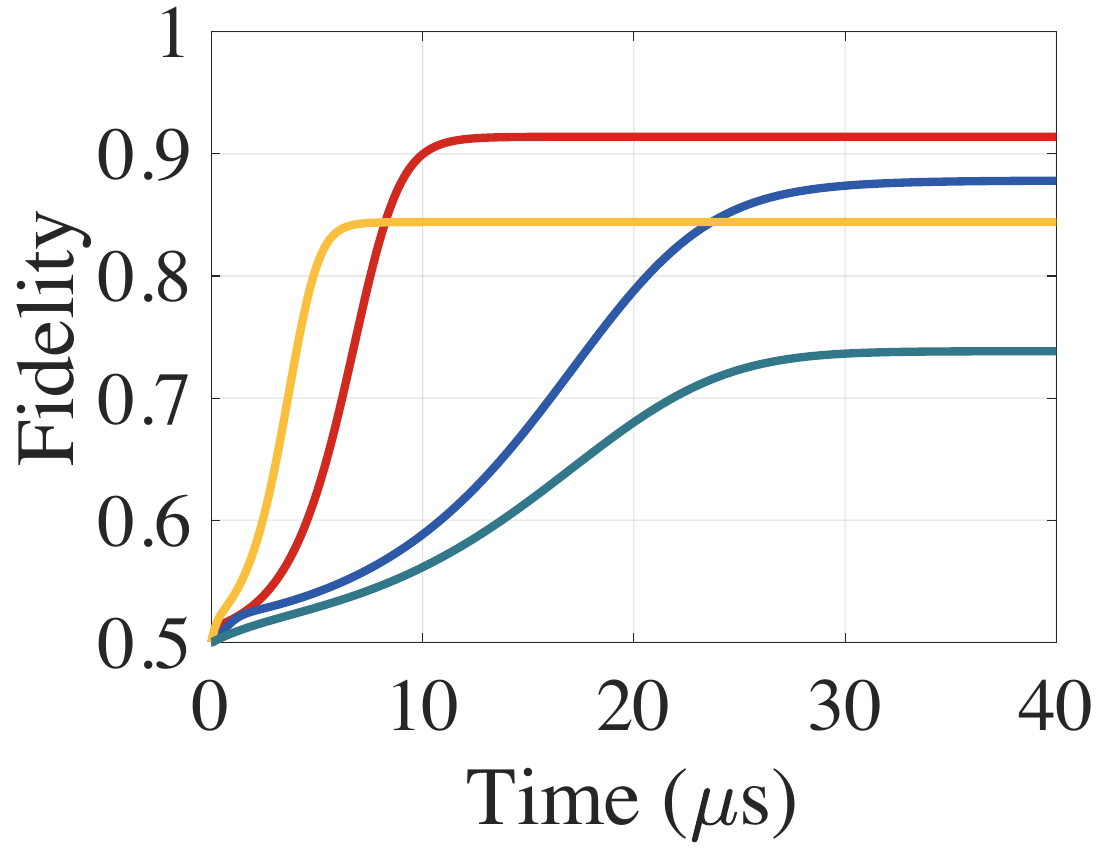}%
  \label{Fig:5a}}
  \hfil
  \subfloat[]{\includegraphics[width=4.3 cm]{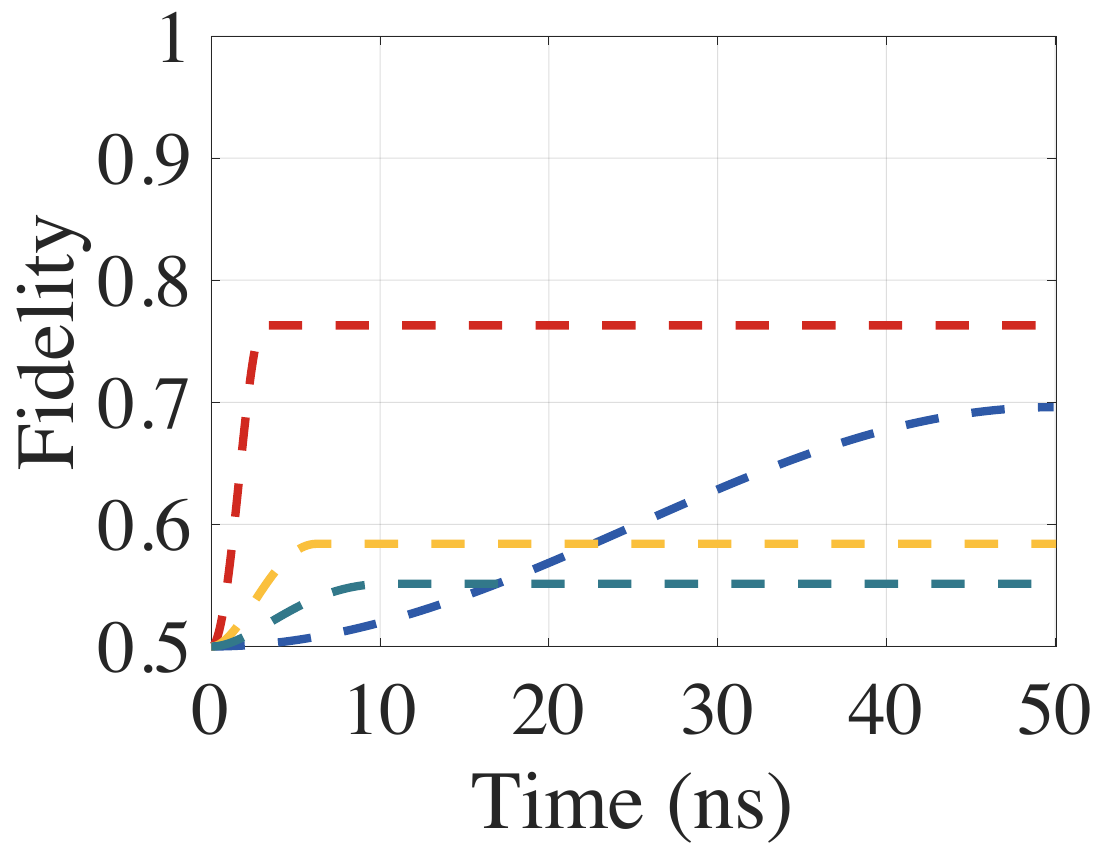}%
  \label{Fig:5b}}
\caption{\label{fig:5} Estimation of random state transmission fidelity between two nodes (a) with STIRAP, and (b) without STIRAP. Red, blue, yellow, and green curves use the state-of-the-art parameter values: $g=$\{100, 38, 98, 21\}$\times 2\pi\ \rm{MHz}$, $\kappa=$\{6, 1.3, 253, 10\}$\times 2\pi\ \rm{MHz}$, and $\gamma=$\{65, 96, 6, 30\}$\times 2\pi\ \rm{MHz}$ from the literature~\cite{young2022architecture,bonizzoni2017coherent,PhysRevLett.105.140501,doi:10.1126/science.aau4691}.}
\end{figure}

Fig.~\ref{Fig:5a} shows the fidelity of a random quantum state transmission between two nodes with STIRAP considering different system parameters. The fidelity rate reaches stability after a certain period of time. Additionally, by comparing Fig.~\ref{Fig:5a} and Fig.~\ref{Fig:5b}, we find that the STIRAP method can significantly improve the fidelity of state transfer, although the effect is different for different systems. However, STIRAP induces a high latency, which calls for a necessary trade-off to achieve high-fidelity state transfer within reasonable delays, particularly given the limited qubit coherence times. 

\subsection{State Transmission over Multiple Hops}
In the red line of Fig.~\ref{Fig:5a}, the fidelity of the transmitted state can be consider stable after 20 $\upmu$s. Hence, we assume that the coupling switch of the transmission operation can be turned off, and the coupling switch of the next transmission (i.e. hop) can be turned after 20 $\upmu$s. As previously noted, this delay can be specified according to system parameters and fidelity requirements. Fig.~\ref{Fig:5a} shows that state fidelity gradually decreases with every state hop, as the initial state becomes a mixed state coupled to the environment.

\subsection{Comparison between Communication Media}
\begin{figure}[!t]
  \centering
  \subfloat[]{\includegraphics[width=4.3 cm]{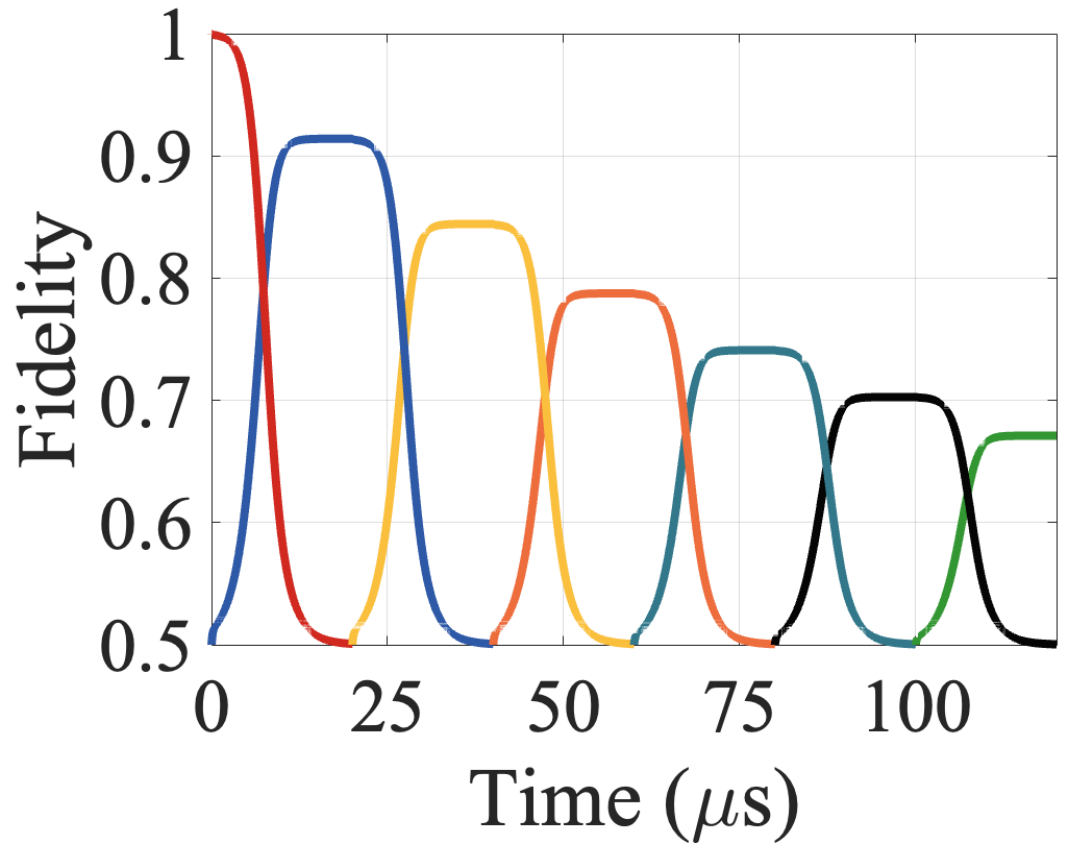}%
  \label{Fig:6a}}
  \hfil
  \subfloat[]{\includegraphics[width=4.3 cm]{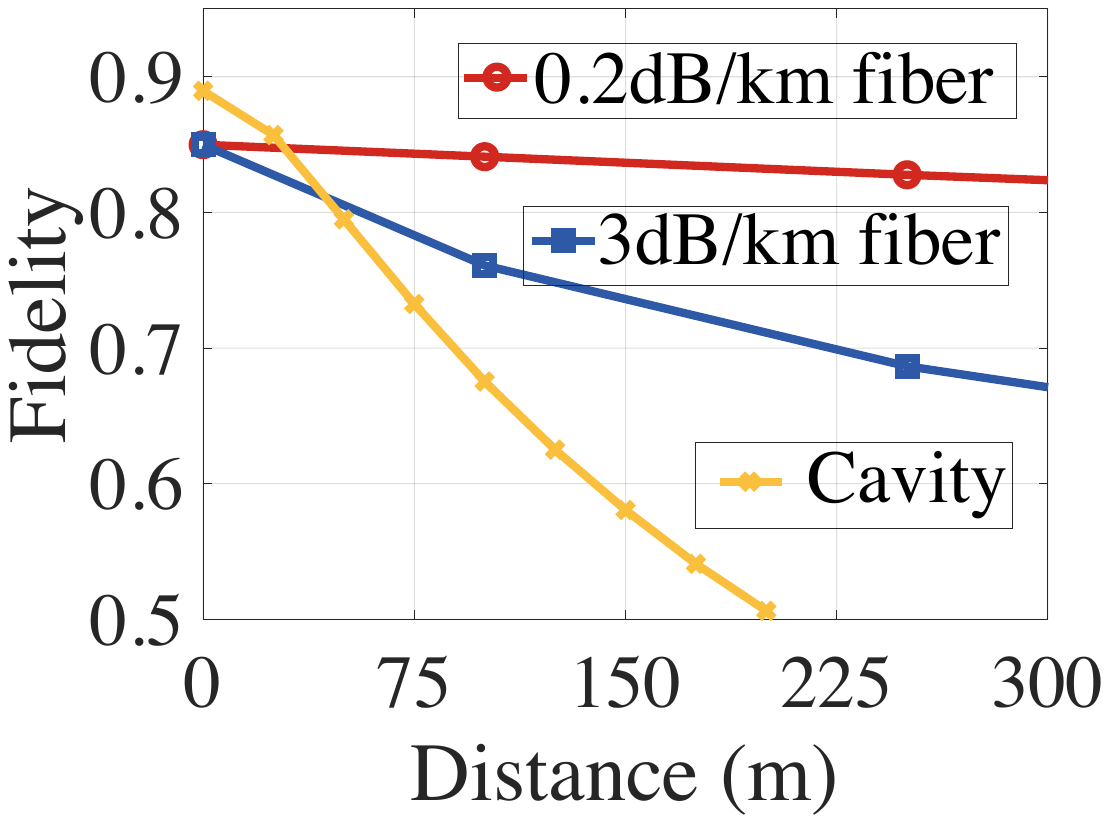}%
  \label{Fig:6b}}
  \caption{\label{fig:6} (a) Estimation of a random state transmission fidelity with STIRAP over a 7-node network with $g_0=100 \times 2\pi\ \rm{MHz}$, $\kappa=6 \times 2\pi\ \rm{MHz}$, and $\gamma=65 \times 2\pi\ \rm{MHz}$. From left to right, the different colors in both figures represent the quantum state fidelity in subsequent nodes, from 1 to 7. (b) Relationship between fidelity and fiber distance when using only the cavity and when using both the cavity and the fiber simultaneously.  }
\end{figure}

Fig.~\ref{Fig:6b} shows the correlation between state transmission fidelity and distance separating nodes. We consider two setups: one using only the cavity as communication medium, and the other where qubits within cavities are coupled to the optical fiber. Considering the same system parameters, the fidelity of state transmission using only cavity-mediated transmission remains advantageous for short distance transmission. For long-distance communication, the fiber loss is much smaller than the cavity loss, yielding an infrastructure of optical fibers more suitable for high-fidelity long range communications.

\section{Discussion}
\label{sec:discussion}
We suggest that cavity-based interconnects are ideal for short-range, cryogenic modular setups, while optical fibers are suited for longer-range communications, such as quantum internet networks, for establishing distributed quantum computing systems. When performing quantum state transmission on an actual network, a better approach is to use methods like STIRAP to improve the fidelity of state transfer.

Although exact analytical models provide benchmarks for ideal performance, they become impractical as system complexity increases and fail to fully capture the dynamics of multi-component quantum networks. To address this, it is required to develop accurate modeling and simulation tools that would enable system-level evaluation of modular quantum network architectures, integrating hardware-aware noise models, and faithfully representing qubit-cavity dynamics. This would bridge the gap between analytical modeling and scalable simulations, thus supporting design-oriented analyses with flexibility and simplicity, as well as accurate performance benchmarking, while maintaining a faithful abstraction of physical behaviors.
\section{Conclusion}
\label{sec:conclusion}
As modular and distributed quantum computing systems move toward large-scale experimental realization, further progress is needed in network characterization, workload optimization, and the development of efficient communication protocols. In this context, we posit that the proposed evaluation framework can play an important role for early-stage exploration of design of such systems.

\bibliographystyle{IEEEtran}
\bibliography{IEEEabrv,./conf.bib}

\end{document}